\documentclass[a4paper,
               refpage       
               keeplastbox,   
               nospread,     
               ]{jacow}
\usepackage{pdfpages,multirow,ragged2e} %
\makeatletter%
	\ifboolexpr{bool{xetex}}
	 {\renewcommand{\Gin@extensions}{.pdf,%
	                    .png,.jpg,.bmp,.pict,.tif,.psd,.mac,.sga,.tga,.gif,%
	                    .eps,.ps,%
	                    }}{}
\makeatother

\newcommand{\be}{\begin{eqnarray}}
\newcommand{\ee}{\end{eqnarray}}

\newcommand{\benum}{\begin{enumerate}}
\newcommand{\eenum}{\end{enumerate}}
\newcommand{\bi}{\begin{itemize}}
\newcommand{\ei}{\end{itemize}}

\usepackage{subcaption}
\usepackage{multicol,lipsum}
\usepackage{mathtools, cuted}

\allowdisplaybreaks[1]

\begin{document}
	
	\title{A primary electron beam facility at CERN}
	
	\author{ ,\\CERN, Geneva, Switzerland}
	\author{T.~{\AA}kesson\textsuperscript{1}, R.~Corsini\textsuperscript{2}, Y.~Dutheil\textsuperscript{2}, L.~Evans\textsuperscript{2}, B.~Goddard\textsuperscript{2}, A.~Grudiev\textsuperscript{2}, A.~Latina\textsuperscript{2}, \\ Y.~Papaphilippou\textsuperscript{2}, S.~Stapnes\textsuperscript{2} \\
		\textsuperscript{1}Lund University, Department of Physics, 221 00 Lund, Sweden \\ \textsuperscript{2}CERN, CH-1211 Geneva 23, Switzerland }

	\maketitle

	\begin{abstract}
		This paper describes the concept of a primary electron beam facility at CERN, to be used for dark gauge force and light dark matter searches. The electron beam is produced in three stages: A Linac accelerates electrons from a photo-cathode up to 3.5 GeV. This beam is injected into the Super Proton Synchrotron, SPS, and accelerated up to a maximum energy of 16 GeV. Finally, the accelerated beam is slowly extracted to an experiment, possibly followed by a fast dump of the remaining electrons to another beamline. The beam parameters are optimized using the requirements of the Light Dark Matter eXperiment (LDMX) as benchmark.
	\end{abstract}

	\section{INTRODUCTION}

\begin{figure}[hb!]
\begin{center}
   \includegraphics[width=8cm]{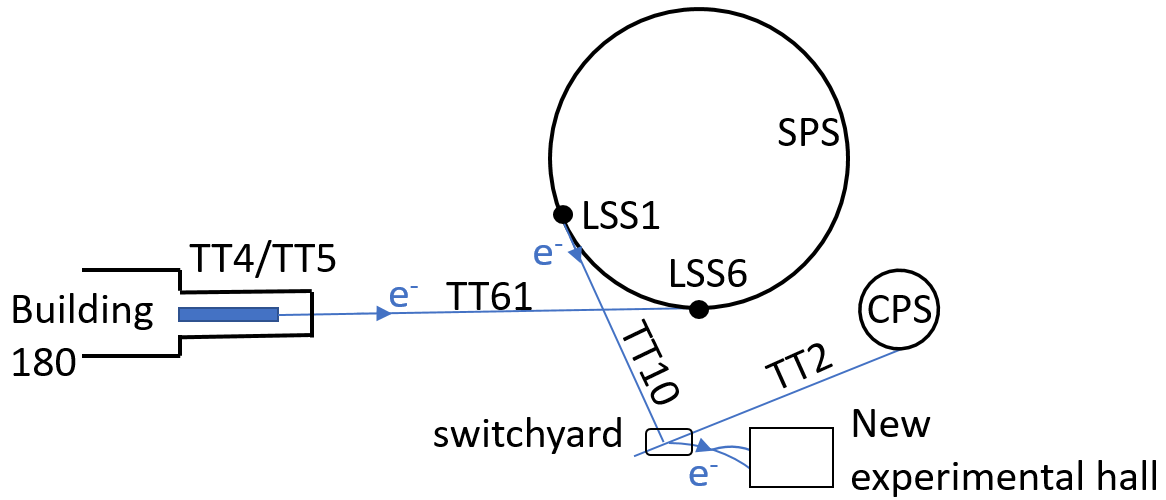}
   \caption{Schematic of the primary electron beam facility.}
\label{fig:eSPS_layout}
\end{center}
\end{figure}
The requirement for a very long spill of low intensity GeV electrons feeding an experiment like the Light Dark Matter eXperiment~\cite{LDMX} for light dark matter searches has motivated a recent proposal for providing this beam at CERN~\cite{LDMX-EOI}, through the Super Proton Synchrotron (SPS). 
The SPS was used as the injector to the Large Electron–Positron collider (LEP)~\cite{LepInjectorStudy:1983aa}, accelerating electrons and positrons from 3.5~GeV to 22~GeV, although most of the associated equipment has now been dismantled. It has been also recently proposed as a pre-injector for FCCee~\cite{FCC-CDR,SPS-FCC}. A 3.5~GeV compact high-gradient linac based on Compact LInear Collider (CLIC)~\cite{CLIC} technology would be injecting pulses of 200~ns duration into the SPS, filling the ring at 100~Hz on a 700~msec duration plateau.
The beam would then be accelerated to 16~GeV, limited by the voltage from 12 Radio Frequency (RF) cavities still available from the LEP days (or a new RF system), which would have to be reinstalled. The electrons would then be resonantly extracted on a 10~s flat top, inside a typical SPS supercycle (e.g. of 32~s). 
The extracted beam would be transported along a beamline to an experimental area. A fast extraction from the machine in one revolution (23~{\textmu}s) is also considered, to feed a possible beam dump type experiment. 
The main accelerator components, shown schematically in Fig.~\ref{fig:eSPS_layout}, are described in this paper.

\section{Electron linac}
\label{sec:eSPS:linac}
The linac consists of two parts; the injector which produces the electron bunches with the required time structure, emittance and charge, and the X-band linac which accelerates them at high-gradient from 150~MeV at the injector output to 3.5~GeV. The beam consists of 40 bunches per pulse with a separation of 5~ns. The bunch spacing corresponds to the RF bucket spacing of the 200~MHz SPS RF system and the number of bunches to the optimized linac pulse length. A 100~Hz linac repetition rate means that the SPS can be filled with 3000~bunches, the maximum given the circumference of the ring and the injection kicker considered, within 1~s. 

\begin{table}[h]
\begin{center}
\caption{Beam Parameters at the End of the CLEAR Injector}
\label{tab:CLEARinjectorPars}
\begin{tabular}{lc}
\toprule
\textbf{Parameter} [unit]   & \textbf{CLEAR range}  / \textbf{LDMX}\\
\midrule
Energy [MeV]  				& 50 to 220           / 150 \\
Bunch charge [nC]           & 0.001 to 1.5        / 0.016 \\
Norm. emittance [{\textmu}m]    & $\sim$3 for 0.05 nC/bunch / 3 \\
							& $\sim$20 for 0.4 nC/bunch     \\
Bunch length rms [mm]     	& 0.3 to 1.2 			/ 0.8 \\
Energy spread rms [\%]    	& below 0.2 			/ 0.1 \\
Number of bunches     		& 1 to 200             	/  40 \\
Micro-bunch spacing [ns]	& multiple of 0.75		/ 5 \\
\bottomrule
\end{tabular}
\end{center}
\vspace{-10pt}
\end{table}

The CERN Linear Electron Accelerator for Research (CLEAR) injector operating at CERN can be used basically unchanged to produce the electron beam with the specifications required for the LDMX and beam dump experiments, as well as for many other potential applications~\cite{GAMBA2017}. 
The range of beam parameters which can be obtained at the end of the CLEAR injector are summarised in Table~\ref{tab:CLEARinjectorPars}. 
Whereas in the injector the S-band (3~GHz) RF structures are used at a relatively low accelerating gradient of about 15~MV/m in order to obtain the desired beam parameters, high gradient X-band (12~GHz) RF accelerating structures are used in the linac in order to make it compact and accelerate the beam from 0.15~GeV up to 3.5~GeV within 70~m. A high gradient is required due to limited space in the Transfer Tunnel 4 and 5 (TT4 and TT5) area. (see Fig.~\ref{fig:eSPS_layout}) The X-band high gradient RF technology has been developed in the framework of the CLIC study and now is being widely adopted. 
For the klystron based option of the first stage of CLIC at 380~GeV,  an average loaded acceleration gradient of 75~MeV/m (95~MeV/m unloaded) has been chosen as a compromise between making the main linac as short as possible and reducing the required peak power and associated number of klystrons. About one klystron per meter is necessary to feed the klystron-based CLIC main linac. Going to slightly lower gradient  reduces the peak power requirement per meter and simplifies the integration in the available space in TT4/TT5 area. The design of the linac is similar to the one of the EuSPARC facility at Frascati~\cite{Diomede:IPAC18-THPMK058} which has an 84~MeV/m average accelerating gradient and for single bunch operation. The longer RF pulse length of 200~ns, needed for trains, results in a reduced gradient of 66~MeV/m because less pulse compression can be used. The parameters of the X-band linac are summarized in Table~\ref{tab:XbandLinacPars}.

\begin{table}[!h]
\begin{center}
\caption{Parameters of the X-band Linac for the LDMX Train, With the Single Bunch Values in Parentheses}
\label{tab:XbandLinacPars}
\begin{tabular}{lc}
\toprule
Parameter  [unit]        	& Train (Single bunch)\\
\midrule
Bunch train length [ns]			& 200   \\
RF pulse length [ns]			& 300 (100)  \\
Energy gain / structure [MeV] 	& 33 (42)  \\
Energy gain / RF unit [MeV] 	& 264 (336)  \\
Tot.~energy gain on crest [MeV]	& 3432 (4368) \\
Pulse compression power gain	& 3 (5) \\
RF units    	       			&          				 13 \\
Filling factor [\%]   	       	&          				 75 \\
Total length  [m]   	       	&          				 70 \\
Number of klystrons		     	&  						 26 \\
Klystron peak power [MW]   		&  						 50 \\
Klystron pulse length [{\textmu}s]	&         				 1.6 \\
\bottomrule
\end{tabular}
\end{center}
\vspace{-10pt}
\end{table}

The linac optics is based on a 90$^\circ$ FODO lattice for beam stability. The  emittance evolution is determined by the wakefield generated in the X-band structures. For high-charges or long-bunches, BNS damping~\cite{Balakin}             must be used to limit emittance growth and prevent beam break-up due to transverse short-range wakefields. For a bunch charge of 1~nC, for example, an acceleration off-crest by 20$^\circ$ effectively reduces the wakefield-induced emittance growth.
The effects of the short-range wakefields on the bunch determines the permitted combinations of bunch charge and lengths that can be transported though the linac. In the presence of short-range wakefields emittance growth and amplification of injection offsets can occur from misalignments of the accelerating structures and off-center injection. 
The threshold of acceptability depends on the beam quality requirements at the end of the linac.
The natural suppression of the long range wakefields, due to the tapering of the iris aperture, also guarantees beam stability in multi-bunch operation for all 40 bunches in the nominal 200~ns long train. Bunch charges up to 300 pC can be transported through the linac without incurring in multi-bunch beam breakup instability. Semi-analytical estimations show that, even with no damping (small $Q$), bunches with that charge are transported with an average amplification factor of the incoming beam injection offsets of less than 2.



\section{Electron beam in the SPS }
\label{sec:eSPS:SPS}


\subsection{Injection to the SPS}
\begin{figure}[h!]
\begin{center}
   \includegraphics[width=8cm]{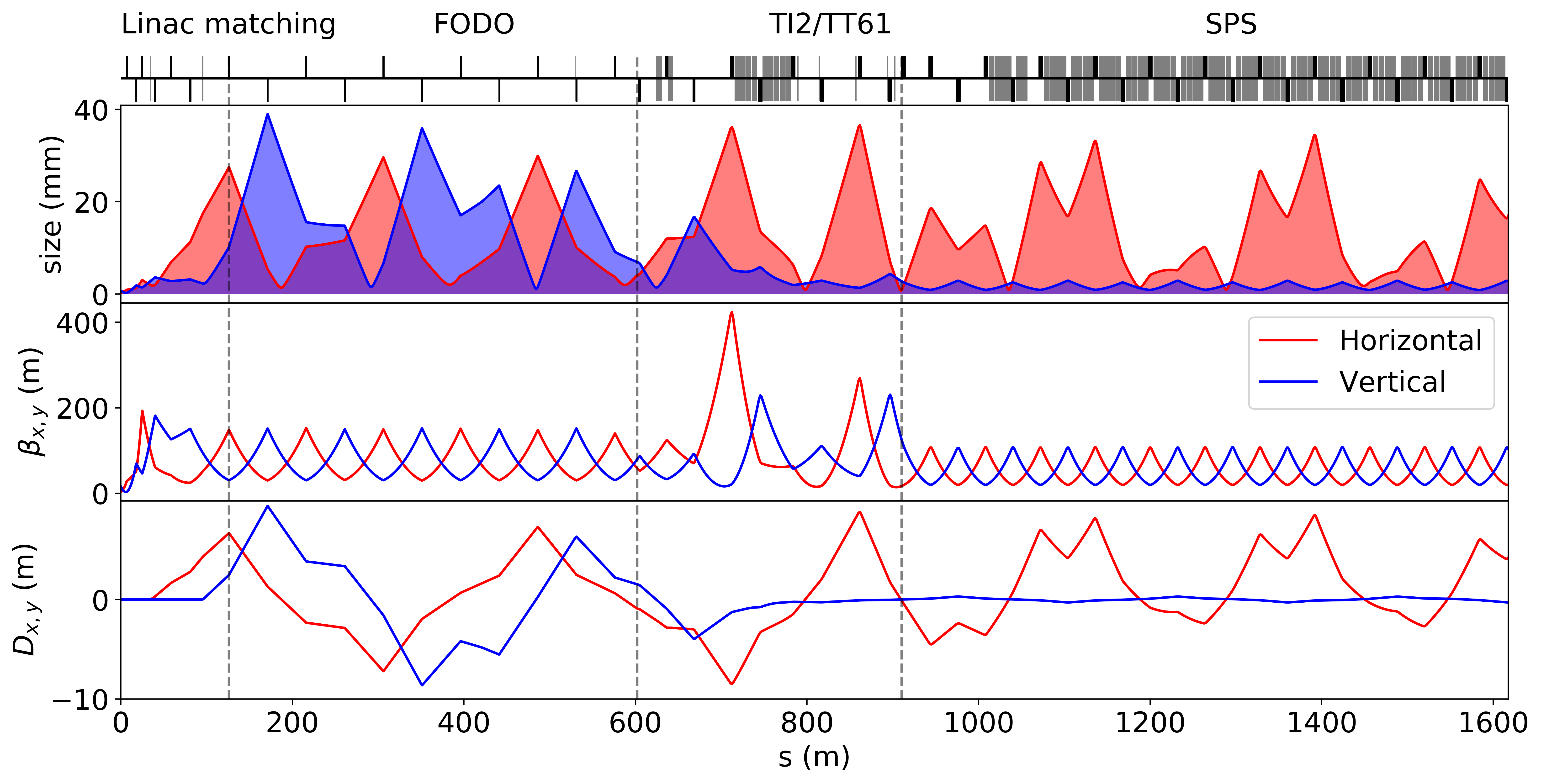}
\\
   \includegraphics[width=8cm]{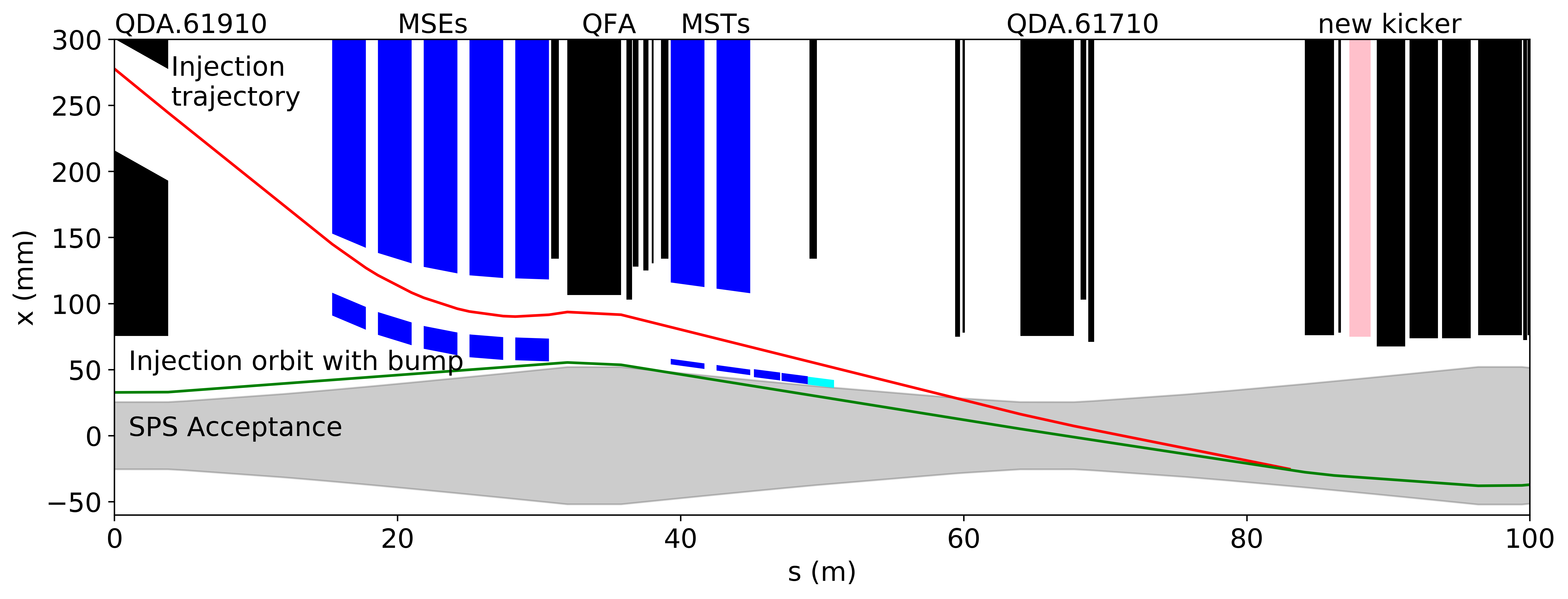}
   \caption{Horizontal (blue) and vertical (red) beam sizes and optical functions from the Linac to the SPS (top),  and horizontal electron injection trajectory in the LSS6 region (bottom).}
\label{fig:Linac_to_SPS}
\end{center}
\vspace{-10pt}
\end{figure}

Electrons from the Linac in TT4 would be transported to the injection system in SPS LSS6. Fig.~\ref{fig:Linac_to_SPS} (top) presents a preliminary design of this transfer line. Beam size is represented using the 4~$\sigma$ envelope with the beam parameters mentioned Table~\ref{tab:CLEARinjectorPars} and includes maximum trajectory offsets of 2~mm. The linac matching section is composed of 6 independently powered quadrupoles. Bending of the trajectory in both horizontal and vertical planes is needed to follow the existing TT61. A long FODO straight section uses low-field air-cooled quadrupoles in a regular arrangement and powered in series. All magnets are available at CERN and would require at most a refurbishment.

From the end of the FODO section the beam joins the TI2 and TT61 lines currently used for transfer of the proton beam from the SPS to the LHC. 
Apertures in this region are limited and will need to be studied in detail, including the beam matching in the SPS. 
The linac beam structure with 200~ns trains of 40 bunches separated by 5~ns calls for a bunch-to-bucket fast injection scheme, as used during the LEP era~\cite{LepInjectorStudy:1983aa}. Fig.~\ref{fig:Linac_to_SPS} (bottom) shows the injection trajectory through the side channel of the quadrupole QDA.61910 followed by the MSE septa bending and deflection onto the SPS orbit by a new kicker. As during LEP operation, the MSE will be powered at low field using a new low current power supply and the MST will be powered-off but degaussed using another power supply.
The 200~ns linac trains require a kicker system with fast rise- or fall-time, to allow maximum filling of the ring. A transmission line of magnets composed of two 50~cm tanks could achieve a rise time of 100~ns and the desired flat-top of 200~ns to provide the required kick of 500~{\textmu}rad. The use of solid state switches would allow to follow the linac 100~Hz repetition rate.
A new kicker system has to be designed and built but will largely make use of existing technologies.


\subsection{SPS Beam Dynamics}

The electron bunches will be accelerated to 16~GeV with an RF system with 12 LEP cavities providing 1~MV RF voltage~\cite{200MHzSPSRFpower:1985,Faugeras:PAC87,Faugeras:PAC89} and can be reinstalled in LSS3. Most of the HOM, fundamental power couplers and tuners have to be rebuilt, in addition to a new high power RF system, including 60~kW power amplifiers for each cavity, with the new LLRF system. 
The RF system for lepton acceleration was removed from the SPS, along with many other items, after LEP decommissioning in order to reduce the impedance, required for the LHC high intensity proton beam. A preliminary estimate for the increase of the longitudinal broad-band impedance, so-called Z/n, is found to be about 0.55 Ohm, based on the values from~\cite{SPSImpedance:1996}. Recent impedance considerations, suggest alternatively  to install a new super-conducting RF system~\cite{ElenaSPSImp}. 


The injection time of around 0.8~s is very short compared to the long damping time which corresponds to 9~s at 3.5~GeV, 
This means that all emittances will essentially preserve their injected characteristics during this short injection plateau.
Particular attention has to be given for the first turn beam threading and establishing the closed orbit. The SPS dipoles have to operate at a very low current of around 40~A, as compared to 6~kA at 450~GeV. During the LEP era, special control loops had to be included in the SPS power convertors in order to guarantee the required current stability~\cite{Cornelis:EPAC88,Cornelis:PAC89}. The SPS is a 6-fold symmetric ring, based on a FODO lattice, with a phase advances of $\pi/2$, providing an integer tune of 26 (Q26-optics).
This was the working point used for injecting leptons in the SPS for LEP. The quadrupoles at injection may have similar low current issues for controlling the optics of the ring. Their strength could be increased by moving the cell horizontal phase advance to around 135$^\circ$ (i.e. $3\pi/4$), which approaches the optimal phase advance for emittance minimisation in FODO cells, 
while keeping the total arc phase advance a multiple of 2$\pi$. This is not strictly necessary for the present application but can open the way to low emittance ring R\&D~\cite{Papaphilippou:IPAC13-TUPME042}.

In particular at injection but also through the cycle (although less critical due to the increase of radiation damping),  it is necessary to mitigate collective instabilities. The classical "head-tail" instability can be suppressed by imposing positive chromaticity, primarily using the SPS sextupoles (two families), throughout the cycle. For establishing a good chromaticity control, effects such as remnant fields in the main dipoles and sextupoles and eddy currents have to be taken into account. Regarding the Transverse Mode Coupling Instability, and based on a broad-band resonator model for a transverse impedance of 10~MV/m~\cite{DBr:PAC89} and typical SPS parameters, the threshold at injection is found at around 2$\times$10$^{10}$ e/bunch,  for a bunch length of 1~mm and RF voltage of 1~MV. 
Some emittance blow-up may be observed due to intrabeam scattering at injection, but this will heavily depend on the injected beam parameters, whose brightness can be tuned to be relatively low, in particular by injecting large transverse and longitudinal emittances. Ion instabilities can be mitigated using trains with gaps and standard multi-bunch transverse feedback systems, as in modern electron storage rings.
The acceleration cycle of 0.2~s although quite short, is below the maximum ramp rate of 90~GeV/s achieved for the lepton cycles in the SPS. At 16~GeV, the dynamics is dominated by synchrotron radiation damping, with transverse damping times of around 94~ms and energy loss per turn of 7.8~MeV. 
The average synchrotron radiation power is below 1~W/m, for the typical beam parameters and considering a duty cycle of 20~\%, which most likely will not necessitate any dedicated radiation absorption scheme.



%
\subsection{Extraction from the SPS}


\begin{figure}[h!]
  \centering
   \includegraphics[height=3.cm]{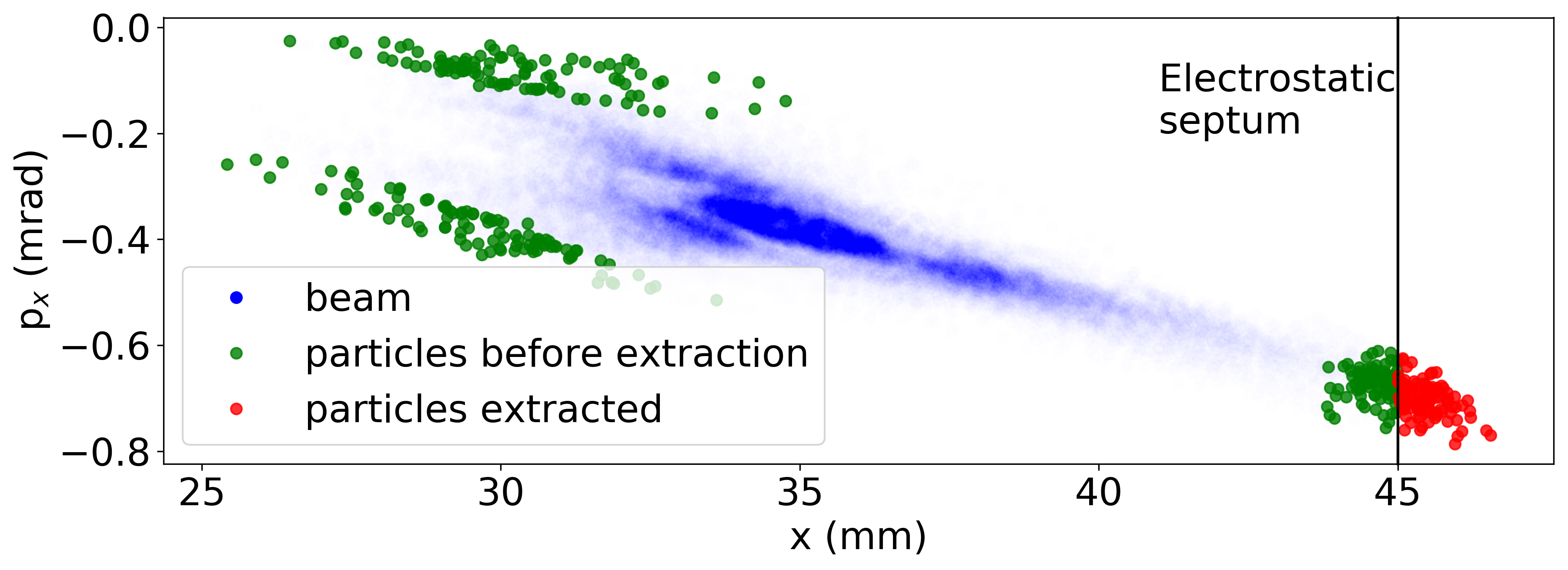}

  \caption{Extraction process in the horizontal phase space at the electrostatic septum.}
  \label{fig:extraction}
\vspace{-10pt}
\end{figure}

Following the experiment requirements, the beam can be extracted by using quantum excitation and amplitude extraction.
Figure~\ref{fig:extraction} shows the phase space simulated using MADX thin tracking. The core of the beam is stable while large amplitude particles are trapped on the third order resonance shown by the triangular beam shape. An electrostatic septum placed at a specific distance from the  core channels particles reaching an excursion of more than 45~mm towards the extraction channel. The extracted beam shown in red on Fig.~\ref{fig:extraction} has a geometrical emittance of $10^{-8}$\,m.rad. Extraction trajectories were investigated in the LSS1 of the SPS, used for injection of the proton beam coming from the Proton Synchrotron (PS). The scheme 
requires a new electrostatic septum of length 2.5~m and field 5~MV/m to provide a deflection of the extracted particles of around 780~{\textmu}rad. Upstream from the existing MSIs, a new thin magnetic septum with a blade of 10~mm and a field of 150~mT will provide an additional deviation of 4~mrad, to reach the existing MSI septa and enter the TT10 line,
used to transport protons from the PS to the SPS. Same magnetic polarity and similar fields to the ones of proton operations can be used for the transport of the extracted electron beam in the opposite direction. 
The first part of the line is powered by only two power supplies for a regular FODO lattice, designed to match the injected proton beam into the SPS lattice. In the case of slow extraction the beam parameters are very different from the ring ones. Therefore the extracted beam is not matched to the FODO lattice of the TT10 line. However transporting the beam in this mismatched transfer line is still possible.

\end{document}